\documentclass[11pt, oneside]{article}

\usepackage{caption}
\usepackage{graphicx}
\usepackage{bm}
\usepackage{geometry}

\title{Absorption of a Particle by a Rotating Black Hole: The Potential Barrier}

\author{Leon Heller
\\Applied Modern Physics Group, P-21, Los Alamos
National Laboratory
\\Los Alamos, New Mexico, 87545
\\LANL Report No. LA-UR-16-23309
\\lheller@lanl.gov}

\begin{document}

\maketitle

\begin{abstract}

For a test particle approaching a rapidly rotating black hole we find a range of values of the particle's energy and angular momentum, on the order of 1\% or more of the corresponding values of the hole,  such that three conditions are satisfied. 1) The particle can reach the horizon. 2) After absorption the new hole still has a horizon. 3) The area of the new hole is less than the area of the original one, in apparent violation of a theorem of Hawking \cite{Hawking}. We offer support for the claim that the test particle approximation is the cause of the violation.


\end{abstract}

\section{Introduction}

In a previous paper on this topic \cite{Heller} a straightforward calculation was made  of the area of a rotating black hole of mass $M$ and angular momentum $J$ before and after absorbing a particle with energy $E$ and angular momentum $L$. Completely neglecting the gravity produced by the particle, i.e., making the test particle approximation, the metric becomes that of Kerr; see \cite{Boyer}. With $J$  near the Kerr limit of $M^2$, it was found that the final area could be smaller than the initial area, which would be a violation of a theorem of Hawking \cite{Hawking}. However, J.  Beckenstein \cite{Beckenstein} asked whether the particles in question could actually reach the horizon if they are coming from a distance, and in the present paper we examine this question. 

\section{Development}

Restricting the particle's motion to lie in the equatorial plane of a black hole with mass $M$ and  angular momentum $J$, the invariant line element for the Kerr metric  expressed  in terms of the Boyer-Lindquist coordinates is  \cite{Boyer}

\begin{equation}
ds^2=g_{\mu \nu}dx^\mu dx^\nu=-(1-\frac{2M}{r})dt^2 -\frac{4Ma}{r}dtd\phi +\frac{r^2}{\Delta}dr^2 +R^2 d\phi ^2
\label{eq:ds^2}
\end{equation}
where $a=J/M$,  $\Delta=r^2-2Mr+a^2$ and $R^2=r^2+a^2+2Ma^2/r$. $\Delta$ vanishes at the horizon. We use units in which Newton's gravitational constant and the speed of light are both set to unity.

The equation $p^2=g^{\mu \nu} p_\mu p_\nu = -m^2$ is quadratic in the particle's energy $E=-p_t$, angular momentum $L=p_\phi$, radial momentum $p_r=g_{rr}p^r$,  and rest mass $m$ for arbitrary values of the radial coordinate $r$. Multiplying through by -$r^2 \Delta$ it is shown in Misner et al.  \cite{Misner} that the equation can be written   

\begin{equation}
\alpha E^2 -2\beta E + \gamma -r^4(p^r)^2 =0
\label{eq:p^2=-m^2}
\end{equation} 
where 
\begin{equation}
\alpha =(r^2+a^2)^2 -\Delta a^2, 
\label{eq:alpha}
\end{equation}

\begin{equation}
\beta = 2 JrL,
\label{eq:beta}
\end{equation}

and
\begin{equation}
\gamma = (2Mr-r^2)L^2 -m^2r^2 \Delta.
\label{eq:gamma}
\end{equation}
The solution of the quadratic equation (\ref{eq:p^2=-m^2}) gives the energy $E$ as a function of $M$, $J$, $r$  $L$, $p^r$ and $m$ \cite{Misner}
\begin{equation}
E=\frac{\beta+  \sqrt{\beta^2 -\alpha \gamma +\alpha r^4 (p^r)^2}}{\alpha}.
\label{eq:E(alpha,beta,gamma,p^r)}
\end{equation}   

It is convenient to introduce a dimensionless variable $u=J/M^2=a/M$, which ranges from zero for a Schwarzschild hole to unity for an extreme Kerr hole. The radius of the horizon of the hole is $r_+=MuK(u)$ and its area is $A=8\pi Mr_+$, where the function $K(u)$ is  defined as 

\begin{equation}
K(u)=\frac{1+\sqrt(1-u^2)}{u}.
\label{eq:K(u)}
\end{equation}
When Eq.(\ref{eq:E(alpha,beta,gamma,p^r)}) is evaluated at the horizon it becomes
\begin{equation}                                     
E= \Omega _H L +\frac{uK(u)}{2} |p^r|
\label{eq:horizonequation}
\end{equation}
where $\Omega _H$ is the angular velocity at the horizon and is equal to $1/(2MK(u))$;  $p^r$ is the value of the radial momentum at the horizon.

\section{Effective Potential, Horizon, and Area}

In this section we examine three issues for a particle with energy $E$ and angular momentum $L$ approaching a black hole from a distance. In $\mathbf{Effective}$  $\mathbf{Potential}$ we ask if the particle can actually reach the horizon.  If it does and is absorbed  , then in $\mathbf {Horizon}$, we test to see if the new hole has a horizon. And finally in $\mathbf{Area}$ we locate the regions in which  the area of the new hole is greater or less than the original area.

\subsection{Effective Potential}

In reference \cite{Misner} the effective potential $V(r)$ is defined as the lowest possible value the particle's energy can have at any given radius $r$, for fixed angular momentum $L$ and mass $m$. This is obtained from Eq.(\ref{eq:E(alpha,beta,gamma,p^r)}) by putting $p^r=0$; it is that value of the energy for which $r$ is a {\it turning point}. The expression for the effective potential is,  therefore
\begin{equation}
V(r)=\frac{\beta+ \sqrt(\beta^2 -\alpha \gamma)}{\alpha}
\label{eq:Veff}
\end{equation}
Just as in Newtonian dynamics the effective potential depends on the particle's angular momentum; in the present case it also depends on the particle's mass.

To test whether a particle arriving from a distance with given energy, angular momentum, and mass can reach the horizon, one plots V as a function of $r$ and looks for its maximum value. If $E$ is less than $V_{max}$ then the particle will reach a turning point and will not make it to the horizon. Furthermore, if a given particle with zero mass can't reach the horizon, then neither can a particle with the same angular momentum but having a non-zero mass. This is seen from Eqs.(\ref{eq:gamma}), (\ref{eq:E(alpha,beta,gamma,p^r)}), and (\ref{eq:Veff}), since $\alpha$  and $\Delta$ are non-negative;  therefore the potential $V$ with non-zero mass is larger than $V$ with zero mass for the same value of $r$.

\begin{figure}[htb]
\includegraphics[width=15 cm]{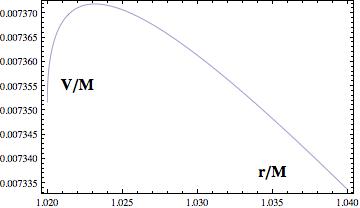}
\caption{A  plot of the effective potential $V$ as a function of the radial coordinate  $r$ for a massless particle with $L/M^2$=0.015 on a black hole with $J/M^2$=0.9998. At the horizon $V/M$ has the value $\Omega _H L/M$, and at infinity the value $m=0$. For this choice of parameters the maximum value of the effective potential is  $V_{max}/M=0.007372$. Any particle having an energy $E/M$ less than this amount will not reach the horizon; and any particle that does reach the horizon will have a non-zero value of the radial momentum there. See Eq.(\ref{eq:horizonequation}).}
\label{fig:V(r)}
\end{figure}

The shape of the potential curve can be seen as follows.  From equations (\ref{eq:alpha}), (\ref{eq:beta}) and (\ref{eq:gamma}) 
 $V(r)$  has the value $\Omega_{H} L$ at the horizon, and the value $m$ at infinity.  Figure \ref{fig:V(r)} shows a typical plot of $V(r)$ choosing $u=J/M^2=0.9998$,  $L/M^2=0.015$, and $m=0$.
 
\subsection{Horizon}

If a particle with energy $E$ and angular momentum $L$  reaches the horizon and is absorbed, it is necessary to know if the new hole still has a horizon. The new values of  of its mass and angular momentum are 

\begin{equation}
M' = M+E
\label{M'}
\end{equation}
and \begin{equation}
J'=J+L.
\label{eq:J'}
\end{equation}
To see if the new parameter $u'=J'/M'^2$ is less than the Kerr limit it is convenient to find the curve in the energy-angular momentum plane on which $u'=1$. This is given by

\begin{equation}
\frac{E}{M}=-1+\sqrt\frac{J+L}{M^2}
\end{equation}

\subsection{Area}

If a particle with energy $E$ and angular momentum $L$ is absorbed by the hole, the ratio of the final area of the hole to the original area is

\begin{equation}
\frac{A'}{A} =\left(\frac{M'}{M}\right)^2 \frac{1+\sqrt(1-u'^2}{1+\sqrt(1-u^2)}.
\label{eq:A'/A}
\end{equation}
We again look for the curve on which  $A'/A=1$.  It takes some algebra to show that the solution for  the particle's energy $E$ as a function of its angular momentum $L$ is

\begin{equation}
\frac{E}{M}=-1+\sqrt(1+\frac{(2JL+L^2)/M^4}{2uK(u)}).
\label{eq:EApreqA}
\end{equation}

 \section{Results}

\begin{figure}[htb]
\includegraphics[width=15cm]{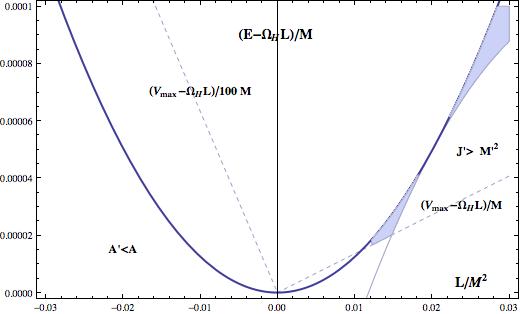}
\caption{The energy-angular momentum plane for a massless particle approaching a rotating black hole having  mass $M$ and angular momentum $J/M^2$=0.9998. As explained in the text, the energy on the ordinate  has been reduced by the product $\Omega_H L$. The final value of the hole's  mass is $M'=M+E$, and its angular momentum  is $J' =J+L$. The dashed curve shows the maximum value of the potential energy barrier $V_{max}$ (also reduced by $\Omega_{H} L$). This is for a massless particle; the two other curves are independent of the particle's mass $m$. On the light solid curve  the final angular momentum $J'$ of the hole is equal to the Kerr limit $M'^2$. For $(E,L)$ combinations {\it above} that curve $J'<M'^2$.
On the dark solid curve the final area of the hole $A'$ is equal to the original area $A$; {\it below} that curve  $A'<A$.      A massless particle with energy and angular momentum lying in the shaded region has the following properties.  (1) It can reach the horizon since its energy $E$ is greater than the maximum of the potential barrier. (2) After absorption it would result in a valid black hole since its angular momentum $J'$ is less than $M'^2$. And  (3) Its final area $A'$ is less than the original area $A$. For a particle with non-zero mass the size of the shaded region is smaller.}
\label{fig:Summary}
\end{figure}

 Figure \ref{fig:Summary} shows the three results obtained in the previous section for a near extreme hole with $u=J/M^2$=0.9998:  the maximum of the effective potential; the curve on which $u'=1$; and the curve on which $A'/A=1$. The abscissa is chosen to be the particle's angular momentum $L$, and the ordinate  is the particle's energy $E$ reduced by the quantity $\Omega_{H}L$. That this combination 
 $E-\Omega_{H} L$ is non-negative  follows from the requirement that the particle's 4-momentum lies in the forward light cone \cite{Misner}. It is, therefore, a {\it constraint} on the allowed values of $E$ and $L$. In reference \cite{Heller} it is shown that the quantity $E-\Omega_{H} L$ is proportional to the absolute value of the particle's radial momentum at the horizon (if it reaches there).
 
 [It so happens that the first variation in the hole's area $\delta A$, with respect to changes in its energy 
 $\delta E$ and angular momentum $\delta L$, is also proportional to the same non-negative quantity 
 $E-\Omega_{H} L$; and this  led some authors to  {\it erroneously} conclude  that the area could never decrease. The error results from the fact that $E$ and $L$ are not completely independent variables.]
 
 In connection with Eq.(\ref{eq:EApreqA}) for the energy that makes $A'/A=1$, we note the expansion in powers of  $L$ of the combination $E- \Omega_{H}L$ 
 
 \begin{equation}
\frac{E-\Omega_H L}{M}=\frac{2 K(u)-1}{8 u K^2(u)} \left(\frac{L}{M^2}\right)^2 +O(L/M^2)^3.
\label{EminusOmegaL}
\end{equation}                 
 For the values of $L$ considered on Fig.\ref{fig:Summary} the quadratic term provides a very good approximation to the numerically exact results plotted there.
 
\section{Discussion}
In reference \cite{Heller}, which also made use of the test particle approximation, no consideration was given as to whether the particle approaching the hole could in fact reach the horizon. From the figure shown there \cite{Heller} it appeared to be possible for a particle with  arbitrarily small values of its energy and angular momentum to violate the area theorem \cite{Hawking}. But those are the very values for which the test particle approximation would be expected to be the most reliable, if  it is reliable at all. Here it is seen that imposing the requirement that the particle's energy be greater than the maximum of the potential barrier, so that it can actually reach the horizon, has considerably reduced the range of energy and angular momentum for which the violation of the area theorem occurs. From Fig.(\ref{fig:Summary}) it is seen that for a hole with $J/M^2$=0.9998, the smallest value of $L/M^2$ for which the particle could reach the horizon, make $u'=J'/M'^2<1$ and $A'<A$ is approximately 0.012, and the corresponding value of $E/M$ is approximately 0.006. 

[In reference \cite{Heller}  there is a figure similar to Fig.(\ref{fig:Summary}) but where the abscissa was chosen to be the particle's energy rather than its angular momentum.  The ordinate,  which was chosen to be the absolute value of the particle's radial momentum at the horizon, $|p^r|$, differs from the ordinate in Fig.\ref{fig:Summary}, $(E-\Omega_{H} L)$, by a factor 
$uK[u]/2$, which is approximately 1/2 for a near extreme Kerr hole. See Eq.(2) in \cite{Heller}.]  

\section{Conclusion}

As compared with the previous paper \cite{Heller}, which also used the test particle approximation but where particles with arbitrarily small values of the energy and angular momentum appeared to be able to violate the area theorem \cite{Hawking}, we have shown here that there are minimum values of the particle's energy and angular momentum, of the order of 1\% of the corresponding values of the hole, such that the particle can reach the horizon and get absorbed, produce a new hole that has a horizon, and still violate the theorem. This makes it  likely that the test particle approximation is the cause of the problem.

To go beyond the test particle approximation requires solving Einstein's equations  taking into account the gravity exerted by the particle on the  hole. Even doing this perturbatively involves a major calculation.

\section{Acknowledgments}

I am indebted to Jacob Beckenstein who asked if  consideration had been given in \cite{Heller}  as to whether the particles that appeared to violate Hawking's area theorem \cite{Hawking} could actually reach the horizon. Since it had not, this led to the calculations in the present paper, and a number of email exchanges  with Professor Beckenstein. After seeing the new results he came independently to the conclusion that the test particle approximation must be the source of the difficulty. His untimely death is both a tremendous loss to the physics community and a severe personal loss for me.




\end{document}